# Engineering Optical and Mirror Bi-stability Mechanically

**Sohail Ahmed[1], Asma Javaid[2], Hui Jing[3], Farhan Saif[1,*]**

[1]*Department of Electronics, Quaid-i-Azam University, Islamabad 45320, Pakistan.*

[2]*Department of Software Engineering, University of Azad Jammu and Kashmir, Muzaffarabad, Pakistan.*

[3]*Key Laboratory of Low-Dimensional Quantum Structures and Quantum Control of Ministry of Education, Department of Physics and Synergetic Innovation Center for Quantum Effects and Applications, Hunan Normal University, Changsha 410081, China.*

[1,*] *Department of Engineering Sciences, University of Electro-Communications, 1-5-1 Chofugaoka, Chofu-shi, Tokyo 182-8585, Japan.*

[*]Corresponding author e-mail: farhan.saif@qau.edu.pk

**Abstract**

We explain optical and mirrors' displacement bistability in a hybrid optomechanical system in the presence of a strong laser driving field and a weak probe field. External modulating fields are applied selectively on movable mirrors. We show that the optomechanical coupling, electro-mechanical Coulomb coupling and, amplitude & phase of external modulating fields are important parameters to control the optical and mirror displacement bi-stable behaviour. The parameters values are taken according to presently available experiments. The study may be applied to the realization of a tunable electro-opto-mechanical switch depending on the optomechanical and Coulomb coupling, frequencies, threshold power, and the amplitude and phase of external mechanical pumps.

## 1 Introduction

In optomechanics the electromagnetic field exerts radiation pressure force that introduces end-mirror movement. This radiation pressure force alters the mean position of the moving mirror, causing the optical path to observe a shift, resulting in a non-linearity of the system. As a back action of the cavity on the field, various non-classical features including optical bistability [1–3] emerges from the induced non-linearity. In their pioneering work, Braginsky and his co-workers [4] showed that radiation pressure force of the confined optical field couples the optical and mechanical modes of a cavity resonator.

In recent years there have been a steady increase of interest in opto-mechanical systems [5], particularly in relation to entanglement of optical modes and mechanical modes [6–10], opto-mechanically induced transparency [11–14], Four-Wave mixing [15, 16] and dynamical localization [17, 18]. Hybrid quantum systems [19–21] have been developed by combining opto-mechanical moving mirror with other systems such as mechanical membranes [22–26], Bose–Einstein condensates/Fermions [26–28], and multi-level atoms [29–33]. Combining opto-mechanical systems with electro-mechanical systems [34–36] make hybrid quantum systems, where the quantum characteristics of the mechanical and the electronic degrees-of-freedom become significant [34–38].

The nonlinear characteristics of the optical cavity field gives rise to optical bistability in the system which exhibits the phenomenon of hysteresis [39]. The optical bistability with Kerr-effect has been investigated in a basic optomechanical system [40,41] as well as in hybrid optomechanical systems composed of trapped cold atoms [42–45] and two-level atoms [59]. Optical bistability has potential applications in nonlinear quantum optics, such as optical signal processing [46], optical switches [47] and optical communication devices [48].

Nano Electro-optomechanical transducers are developed by combining opto-mechanical and an electro-mechanical system [34, 49–56]. These systems are useful to comprehend induced transparency profile [14, 57, 58], super- and sub-luminal effects [26, 27], Fano resonances [11, 59, 60], and Four-Wave Mixing [16, 33, 34]. In this contribution we discuss a nano-electro-opto-mechanical system (NEOMS) which is made up of two charged moving mirrors $MR_1$ and $MR_2$. The optical cavity field is coupled to $MR_1$ via opto-mechanical coupling. Two external biased voltages $+V_1$ and $-V_2$ are applied to the moving mirrors $MR_1$ and $MR_2$, which results a Coulomb coupling between the mirrors. We control optical bi-stability in the probe field and in the mirrors' displacement that takes place in the micro-mechanical motion of the mirrors, $MR_1$ and $MR_2$. We control the bi-stability by controlling the ampli-tude and the phase variations of external modulating fields applied selectively on the moving mirrors. The cavity opto-mechanical system has been realized with a double-micro-disk resonator, a nano-beam photonic crystal, or a microwave device with two micro-mechanical beams [61, 62]. Strong mechanical driving has been utilized to make hybrid quantum spin-phonon devices [63] and ultra-strong exciton-phonon coupling [64].

Optical bi-stability has been investigated in several opto-mechanical systems [40, 65–71] as a result of the dynamic back-action caused by radiation pressure force inside an optical cavity. We further present an adjustable switch based on the controlled bi-stability using various experimental conditions.

The manuscript is organized as follows: In Section 2 we derive the total Hamiltonian of the system and the renowned quantum Langevin equations of motion and obtain steady-state solutions. In section 3, we investigate the bi-stable behaviour of the system's intra-cavity photon numbers in the presence of two external driving fields on the microscopic moving mirrors $MR_1$ and $MR_2$. Moreover, in Section 4 the mirror bi-stability of the steady-state positions of the moving mirrors, $MR_i$, where $i = 1, 2$ in the presence of external modulations is analysed. In Section 5, we conclude our research work and discuss the obtained results.

## 2 The Model

We consider a high-Q Fabry-Pérot cavity of length $L$. The cavity is composed of a fixed mirror $MR_{fixed}$ and a moving mirror $MR_1$, with a driving field $\varepsilon_l$ and a probe field $\varepsilon_p$, as shown in Fig. 1. The mirrors $MR_1$ and $MR_2$ are driven by external modulating fields $\varepsilon_1$ and $\varepsilon_2$, respectively. In addition, $MR_1$ is coupled to the cavity field through radiation pressure force and is coupled to another moving mirror $MR_2$ via a tunable electro-static Coulomb coupling force. The oscillation frequency of $MR_1$ is $\omega_1$ and that of $MR_2$ is $\omega_2$. The Coulomb coupling is controlled by the bias voltages $+V_1$ and $-V_2$ associated to MR$_1$ and MR$_2$, respectively. Hence, in addition to opto-mechanical coupling, the system features a tunable Coulomb coupling strength. The total Hamiltonian of the system is

$$H = H_{mc} + H_{dr} + H_{int}. \tag{1}$$

where $H_{mc}$ is mirror and field Hamiltonian in the absence of any coupling, that is

$$H_{mc} = \hbar\Delta_c c^\dagger c + \left[\frac{p_1^2}{2m_1} + \frac{1}{2}m_1\omega_1^2 q_1^2\right] + \left[\frac{p_2^2}{2m_2} + \frac{1}{2}m_2\omega_2^2 q_2^2\right], \tag{2}$$

under the rotating reference frame at the frequency $\omega_l$. Here, $\Delta_c = \omega_c - \omega_l$ is detuning of the cavity field frequency. The first term represents the single-mode of cavity field with frequency $\omega_c$ and photon annihilation (creation) operator c (c$^\dagger$). The second and third term represent the free Hamiltonian of moving mirrors MR$i$ of mass m$i$ oscillating with frequency $\omega_i$, where i = 1, 2. The operators, q$i$, and p$i$ represent the positions, and momenta, respectively.

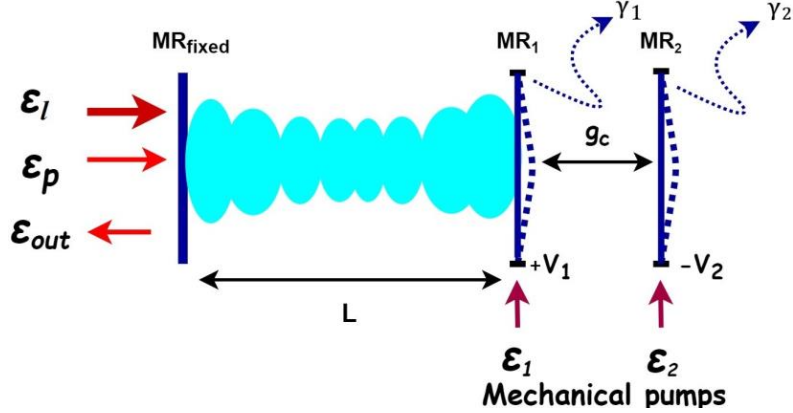

Figure 1: The schematic diagram of the nano-electro-opto-mechanical system: The moving mirror MR$_1$ is coupled with the cavity field through optomechanical coupling $g_0$ and to the second moving mirror MR$_2$ through the coulomb coupling strength $g_c$. A strong input laser field of amplitude $\varepsilon_l$ and a weak probe field of amplitude $\varepsilon_p$ are induced to the cavity through the fixed mirror. In addition, MR$_1$ is charged by the biased voltage $+V_1$, while MR$_2$ is charged by the biased voltage $-V_2$. Here, $L$ is the length of cavity, $\varepsilon_1$ and $\varepsilon_2$ are external driving fields on MR$_1$ and MR$_2$, $\varepsilon_{out}$ is the output probe field and $\gamma_1$ and $\gamma_2$ are mirror decay rates. [56]

The mechanically oscillating moving mirrors MR$_1$ and MR$_2$ can be expressed in terms of phonon ladder operators, such that $q_j = \sqrt{\frac{\hbar}{2m_j\omega_j}}(b_j + b_j^\dagger)$, whereas the corresponding conjugate momenta are $p_j = -i\sqrt{\frac{m_j\hbar\omega_j}{2}}(b_j - \backslash b_j^\dagger)$. Here, j = 1, 2. In terms of the phonon annihilation and creation operators, $bj$ and $b_j^\dagger$, respectively, the Eq. (2) becomes,

$$H_{mc} = \hbar\Delta_c c^\dagger c + \hbar\omega_1 b_1^\dagger b_1 + \hbar\omega_2 b_2^\dagger b_2, \tag{3}$$

we have dropped the constant terms. The interaction Hamiltonian is given by,

$$H_{int} = -\hbar g_0 c^\dagger c q_1 + \hbar g_c q_1 q_2, \tag{4}$$

where, $g_0$ is the opto-mechanical coupling strength, written as $g_0 = \frac{\omega_c}{L}$, and the Coulomb coupling strength $g_c$ is $g_c = \frac{k_e C_1 V_1 C_2 V_2}{\hbar r_0^3}$. Here, $k_e$ is the electrostatic constant while $C_1(C_2)$ and $V_1(V_2)$ are the capacitance and voltage of $MR_1(MR_2)$. The equilibrium spacing between two mirrors is $r_0$.

Moreover, in terms of $b_j$ and $b_j^\dagger$, Eq.(4) is written as,

$$H_{int} = -\hbar G_0 c^\dagger c (b_1 + b_1^\dagger) + \hbar G_c (b_1^\dagger b_2 + b_1 b_2^\dagger). \tag{5}$$

The Hamiltonian $H_{int}$ expresses the opto-mechanical coupling between the cavity field and MR$_1$ and the Coulomb coupling between MR$_1$ and MR$_2$. Here, we obtain $G_0 = g_0\sqrt{\frac{\hbar}{2m_1\omega_1}}$ and $G_c = g_c\sqrt{\frac{1}{m_1 m_2 \omega_1 \omega_2}}$.

Furthermore, $H_{dr}$ is the combined Hamiltonian of the strong laser field of amplitude $\varepsilon_l$, weak probe field of amplitude $\varepsilon_p$, and the external driving fields $\varepsilon_1$ and $\varepsilon_2$. We express

$$H_{dr} = i\hbar\varepsilon_l c^\dagger + i\hbar\varepsilon_p e^{-i\delta t} c^\dagger + i\hbar \sum_{j=1}^{2} \epsilon_j b_j^\dagger e^{-i(\omega_{jf}t+i\varphi_j)} - H.c. \tag{6}$$

where, $\delta = \omega_p - \omega_l$ is the probe field detuning frequency with the laser frequency $\omega_l$ and $\delta_j = \omega_j - \omega_l$.

The classical light fields (pump and probe fields) with frequencies $\omega_l$ and $\omega_p$ are represented in the first two terms. The strong laser power $\wp_l$ and probe field power $\wp_p$ is related to $\omega_l$ and $\omega_p$ by the relations $\varepsilon_l = \sqrt{2\kappa\wp_l/\hbar\omega_l}$ and $\varepsilon_p = \sqrt{2\kappa\wp_p/\hbar\omega_p}$, respectively.

The last term represents the Hamiltonian of external modulating fields on mechanically moving mirror MR$_1$ and MR$_2$ Here, $\epsilon_j$ is the amplitude, $\omega_{jf}$ is the frequency, and $\varphi_j$ is the phase of the driving field $\varepsilon_j$, ($j = 1,2$).

The quantum–Langevin equations for $c, b_1$ and $b_2$ are written as

$$\dot{c} = -\left(i\Delta_c + \frac{\kappa}{2}\right)c + iG_0(b_1^\dagger + b_1)c + \varepsilon_l + \varepsilon_p e^{-i\delta t} + \sqrt{2\kappa}c_{in}(t),$$

$$\dot{b}_1 = -\left(i\omega_1 + \frac{\gamma_1}{2}\right)b_1 + iG_0 c^\dagger c - iG_c b_2 + \epsilon_1 e^{-i(\omega_{1f}t+i\varphi_1)} + \sqrt{2\gamma_1}\xi_1(t),$$

$$\dot{b}_2 = -\left(i\omega_2 + \frac{\gamma_2}{2}\right)b_2 - iG_c b_1 + \epsilon_2 e^{-i(\omega_{2f}t+i\varphi_2)} + \sqrt{2\gamma_2}\xi_2(t) \tag{7}$$

We write $c_{in}(t)$ representing the input vacuum noise with zero mean value associated with cavity field, while terms $\xi_1(t)$ and $\xi_2(t)$ are Brownian noise operators associated with the damping of the MR$_1$ and MR$_2$, respectively. The symbols $\kappa$ and $\gamma_i$ denote the decays terms associated with the cavity and the $MR_i$ ($i$=1,2), respectively. Using mean-field approximation we obtain,

$$\langle\dot{c}\rangle = -\left(i\Delta_c + \frac{\kappa}{2}\right)\langle c\rangle + iG_0(\langle b_1^\dagger\rangle + \langle b_1\rangle)\langle c\rangle + \varepsilon_l + \varepsilon_p e^{-i\delta t},$$

$$\langle \dot{b}_1 \rangle = -\left(i\omega_1 + \frac{\gamma_1}{2}\right)\langle b_1 \rangle + iG_0\langle c^\dagger\rangle\langle c\rangle - iG_c\langle b_2\rangle + \epsilon_1 e^{-i(\omega_{1f}t+i\varphi_1)},$$

$$\langle \dot{b}_2 \rangle = -\left(i\omega_2 + \frac{\gamma_2}{2}\right)\langle b_2 \rangle - iG_c\langle b_1\rangle + \epsilon_2 e^{-i(\omega_{2f}t+i\varphi_2)} \quad (8)$$

The set of equations(8) together with their Hermitian conjugate counterparts ($c^\dagger, b_1^\dagger$ and $b_2^\dagger$) present a set of six coupled first-order differential equations. Such a system may lead to complex dynamics [72–74]. Therefore, an exact solution of these equations is not possible. Hence, we use the ansatz., $\langle \hat{h}\rangle = h_s + h_+ e^{i\delta t} + h_- e^{-i\delta t}$ ,and get the steady-state solutions [75].

Here, $h_s$ denotes any of the steady-state solutions $c_s, b_{1s}$ and $b_{2s}$ and $h_+$ and $h_-$ represent the small perturbations and are of the same order as $\varepsilon_p$. For the case $h_s \gg h_\pm$, substituting the ansatz into Eq. (8), we obtain the following steady-state solutions:

$$c_s = \frac{\varepsilon_l + \varepsilon_p e^{-i\delta t}}{i\Delta + \frac{\kappa}{2}}$$

$$b_{1s} = \frac{iG_0|c_s|^2 - iG_c b_{2s} + \varepsilon_1 e^{-i\phi_1}}{i\omega_1 + \frac{\gamma_1}{2}}$$

$$b_{2s} = \frac{-iG_c b_{1s} + \varepsilon_2 e^{-i\phi_2}}{i\omega_2 + \frac{\gamma_2}{2}} \quad (9)$$

Here, $\Delta = \Delta_c - G_0(b_{1s}^* + b_{1s})$ is the effective detuning, and $\phi_j = \omega_{jf}t + \varphi_j$. Also, $\wp_p \ll \wp_l$, hence, for the simulation only strong laser power is considered.

## 3 Optical Bi-Stability

Here in our scheme bistable behavior is due to the non-linearity which emerges from optomechanical coupling strength $G_0$ as well as Coulomb coupling strength $G_c$. By solving Eq. (9) for the steady state value of cavity field photon number, we have

$$|\varepsilon_l + \varepsilon_p e^{-i\delta t}|^2 = |c_s|^2 \left[\frac{\kappa^2}{4} + \left(\Delta_c - G_0(b_{1s}^* + b_{1s})\right)^2\right], \quad (10)$$

here $|c_s|^2 = c_s c_s^*$. This equation indicates the occurrence of bistable behavior in the system. From Eq. (10) we find that the bistability in photon number vanishes if we have $G_0 = 0$. By rearranging the Eq.(8),
we obtain a third order polynomial of the steady-state intracavity photon numbers as follow:

$$a_1 x^3 + a_2 x^2 + a_3 x + a_4 = 0, \quad (11)$$

where,

$$x = |c_s|^2,$$

$$a_1 = G_0{}^2 \alpha_1{}^2$$

$$a_2 = 2\alpha_1(\Gamma + \Delta_c G_0)$$

$$a_3 = \left(\frac{\kappa^2}{4} + \Delta_c{}^2 + G_0\Gamma(G_0\Gamma + 2\Delta_c)\right)$$

$$a_4 = -|\varepsilon_l + \varepsilon_p e^{-i\delta t}|^2 \qquad (12)$$

Here, $\alpha_1 = \beta_1 + \beta_1^*$, $\Gamma = \alpha_2\varepsilon_2 + \alpha_3\varepsilon_1$, $\alpha_2 = \beta_2 e^{-i\phi_2} + \beta_2^* e^{i\phi_2}$, $\alpha_3 = \beta_3 e^{-i\phi_1} + \beta_3^* e^{i\phi_1}$, where, $\beta_1 = \frac{i(i\omega_2+\gamma_2/2)G_0}{(i\omega_1+\gamma_1/2)(i\omega_2+\gamma_2/2)+G_c{}^2}$, $\beta_2 = -\frac{G_c}{G_0}\frac{\beta_1}{(i\omega_2+\gamma_2/2)}$ and $\beta_3 = \frac{-i\beta_1}{G_0}$.

This equation has three roots, two of which are for stable regimes and the third one is for the unstable regime of the steady-state photon number. The respective branches of the bistable curve have been found by using the solutions of the cubic polynomial equation $ay^3 + by^2 + cy + d = 0$ with inflection and critical points, i.e., $y_c = \frac{-b\pm\sqrt{b^2-3dc}}{3a}$ and $y_{inf} = -b/3a$. Using the solutions for $y_c$ and $y_{inf}$, we can write the critical and inflection points of the Eq. (11) as,

$$x_{c\pm} = \frac{-a_2/a_1 \pm \sqrt{(a_2/a_1)^2 - 3a_3/a_1}}{3}, \qquad x_{inf} = -a_2/3a_1 \qquad (13)$$

where $x_{c+}$ and $x_{c-}$ are critical points of upper and lower stable limbs of the bistable curve respectively, while $x_{inf}$ is the inflection point of the curve. The range of bistability window is determined by these critical points and at these points the driving laser field power $\wp_l$ has a corresponding window also. Our proposed system, we suppose, operates in resolved side-band regime, i.e., $\kappa \ll \omega 1$, and the resolved side-band regime is directly affected by the laser field intensity, hence increasing the laser field power increases the photon number. As a result, for a certain range of strong laser power values, the steady-state photon number shows the bistability. As the driving field intensity is increased, the cavity field detuning approaches a particular value, known as critical detuning. The outset of the bistable behaviour in the system can be seen at this critical value. The critical value of the cavity detuning that determines the condition for the occurrence of bistability in the system, is found by using Eq. (9) as,

$$\Delta_c = \sqrt{3} \qquad (14)$$

To investigate the control of optical and mirror bistability in NEOMS, we consider the parametric values from the recent experiments [29, 47, 48]. The length of the cavity of proposed nano-electro-opto-mechanical system (NEOMS) is taken as L = 25cm and $r_0$ = 2mm. For simplicity, we choose identical moving mirrors (MRs) with masses $m_1(m_2)$= 145ng, oscillation frequencies $\omega_1(\omega_2) = 2\pi \times 947$kHz and decay rates $\gamma_1(\gamma_2) = 2\pi \times 140$kHz. The cavity decay rate is considered as $\kappa = 2\pi \times 215$kHz. The cavity field frequency is calculated by the relation $\omega_c = 2\pi c/\lambda_l$, where $c$ is speed of light, and $\lambda_l$ = 1064nm is the wavelength of the driving field [76]. We choose
$\Delta_c = 3\kappa$ and $\omega_l = \omega_c \omega_1$, and the power of the pump field is taken as 9mW. As the value of $MR_s$ frequency is greater than cavity decay rate $\kappa$, hence the system stays in the resolved side-band regime.

Fig. 2 depicts the bifurcation diagram for the steady-state solution of the Eq. (8), where the steady-state photon number is displayed as a function of the input laser power $\wp_l$ using the value of cavity field detuning $\Delta_c > \sqrt{3}$. The critical points denote the curve's lower and higher stable branches, respectively, whereas the inflection points P and Q denote the curve's unstable branch. When we scan the system with a low laser field and gradually raise the laser power $\wp_l$, the intracavity field photon intensity first follows the lower stable branch $S_1$ of the curve, which then expands to the first critical point $x_{c-}$. The value of steady-state photon number leaps to the upper stable branch $S_2$, which spans from the second critical point $x_{c+}$ to infinity, if the driving laser power is increased to a value of 7.6 mW. The blue-dashed line, which represents the central branch, is unstable and extends from inflection point P to another inflection point Q. The slope of unstable branch is negative; therefore, it cannot be observed empirically. If we start scanning with a larger driving field value and gradually reduce the laser power $\wp_l$, the intracavity photon number will start decreasing by following the upper stable branch initially, however, when it reaches the inflection point P, it will jump to the lower stable branch at second critical point $x_{c-}$ and continue to decrease further.

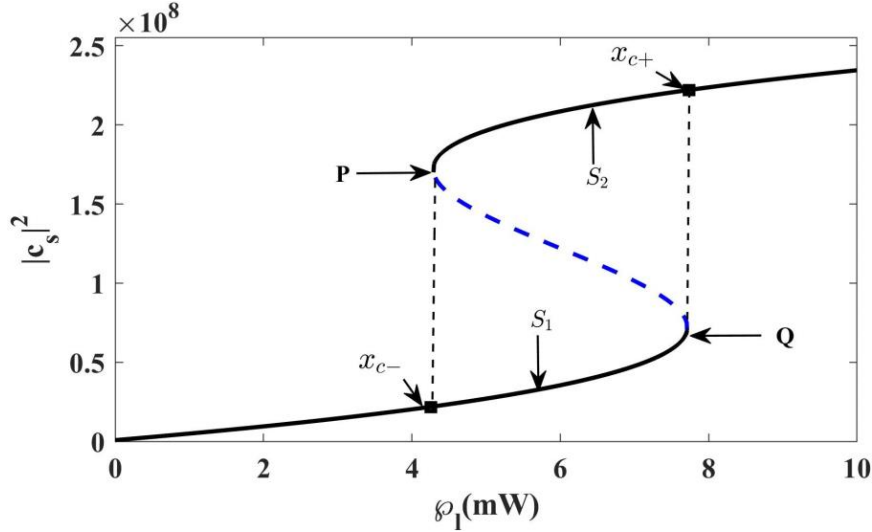

Figure 2: Plot of mean intra-cavity photon number $|c_s|^2$ as a function of the driving laser power $\wp_l$. Other system parameters used for this particular case are $m_1 = m_2 = 145$ ng, $\omega_1 = \omega_2 = 2\pi \times 947$ kHz, $\varepsilon_1 = \varepsilon_2 = 0$, $G_0 = 2\pi \times 5$ kHz, $\Delta_c = 3.6\kappa$ and $\kappa = 2\pi \times 215$ kHz.

We now give a full analysis of controllable bistability, which is primarily influenced by system factors such as coupling frequencies and external modulating field intensity. For various levels of optomechanical coupling strength, the S-shaped bistability curve of intracavity photon counts against laser power $\wp_l$ is shown in Fig. (3). The bifurcation curves overlap for different levels of optomechanical coupling frequency $G_0$, as shown in Fig. (3). As the optomechanical coupling strength $G_0$ increases, the mechanical back-action of the mirror increases the radiation pressure force, causing photon dispersion in the cavity. The bifurcation curve follows the upper stable branch in the lower coupling regime that is $G_0/2\pi = 6$ kHz (black solid line). The lower stable path is followed by the upper stable path as the frequency of optomechanical coupling is increased further. The lower stable path is followed by the upper stable path as the frequency of optomechanical coupling is increased further that is $G_0/2\pi = 7$ kHz (black dashed curve). In addition to the change in photon counts, raising the optomechanical coupling frequency reduces the width of the bistable curve. When the frequency of optomechanical coupling strength $G_0$ increases, the second lower stable path follows the third lower stable path, and vice versa. We see that as the optomechanical frequency changes, the bistability curve changes as well, and the photon number decreases. The flip-flop phenomenon is the name given to this change in photon number. This method provides an elegant way to adjust the intensity of photons in the intracavity. As a result of this control parameter, an experimental realization of controllable optical switch is possible.

The influence of Coulomb coupling strength $G_c$ on optical bistability is now investigated. As a result, we exhibit the intracavity photon number $|c_s|^2$ bifurcation curve as a function of input laser power $\wp_l$ for various values of Coulomb coupling strength $G_c$ in Fig. (4). The bias gates +$V_1$ and -$V_2$ are used to tailor the system for varied Coulomb coupling $G_c$ values. The Coulomb coupling strength i.e., $G_c = 0.2$ MHz (see black solid curve) is not strong enough to change the radiation pressure inside the cavity when the bias gate across each mirror is low. As a result, we have the maximum amount of photons and the system remains in upper stable branch. When the charges on the mirrors are increased by increasing the bias gates +$V_1$ and -$V_2$ across the mirrors MR$_1$ and MR$_2$ respectively, the Coulomb coupling strength is enhanced which, in turn, increases the radiation pressure force inside the optical cavity. Therefore, the nonlinearity in the system is boosted up and as a result the intracavity photon number $|c_s|^2$ is suppressed inside the optical cavity.

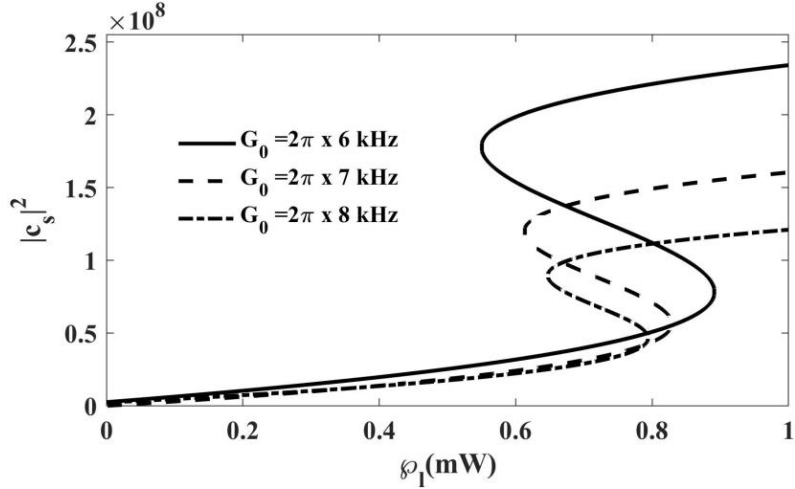

Figure 3: Plot of mean intra-cavity photon number $|c_s|^2$ as a function of the driving laser power $\wp_l$ for different values of optomechanical coupling strength $G_0$. Other system parameters are the same as used in Fig. (2)

This behavior of the system is depicted in Fig. (4) for $G_c = 0.4$ MHz (see red dashed curve) and $G_c = 0.6$ MHz (see blue long-dashed curve). This study reveals that Coulomb coupling $G_c$ has a considerable impact on the bistable behaviour of the steady-state photon number, and that parametric modulation of Coulomb coupling can be used to create a tunable optical switch. Now, we probe the behavior of bistable curve of intracavity photon number $|c_s|^2$ as a function of input laser power $\wp_l$ for different values of cavity detuning $\Delta_c$. For this purpose, we illustrate the effect of the variations in cavity field detuning on the intracavity photon number $|c_s|^2$, in Fig. (5). The stable and unstable points are shifted at different values of photon number by increasing the cavity field detuning frequency. Another prominent effect occurs by increasing $\Delta c$ i.e., the width of bistable curve increases. The bistable characteristic essentially vanishes at decreasing amounts of cavity field detuning i.e, $\Delta c = 1.8\kappa$ (see red solid curve).

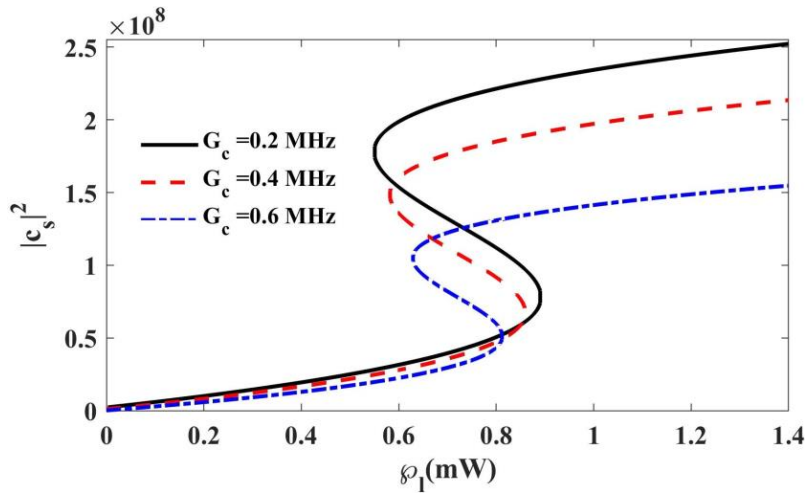

Figure 4: The intra-cavity photon number $|c_s|^2$ as a function of the driving laser power $\wp_l$ is plotted for different values of Coulomb coupling strength $G_c$. Other system parameters are same as used in Fig. (2).

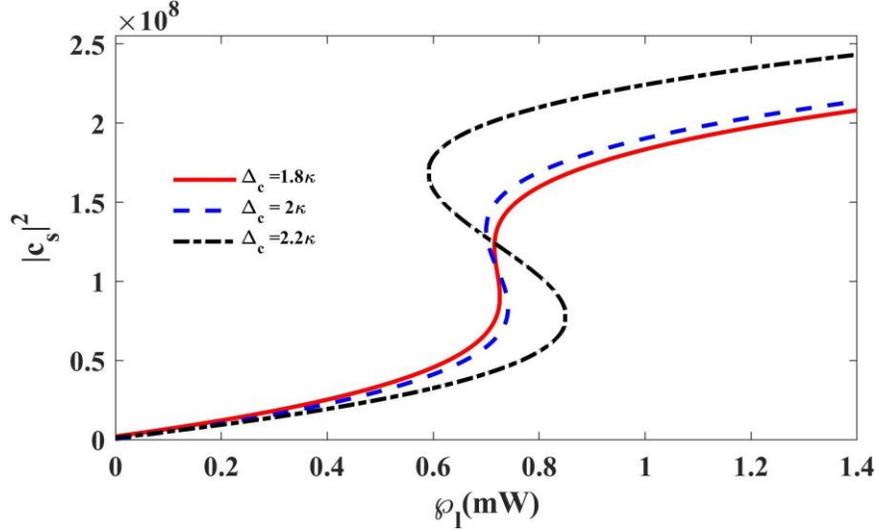

Figure 5: Variation of mean intra-cavity photon number $|c_s|^2$ for bistability as a function of input laser power $\wp_l$ for different values of cavity detuning $\Delta_c$. The remaining parameters are the same as used in Fig. (2).

This is due to the fact that at the critical detuning $\Delta c = \sqrt{3}\kappa$, the system experiences threshold bistability. Higher amounts of cavity field detuning cause the upper stable path of the bifurcation curve to leap to the next upper stable branch (see blue dashed and black dot-dashed curves). A comparable transition from the upper stable path to the lower stable path occurs as the input laser field's power decreases. The genesis of the bistability is shown by these properties, which are the distinctive non-linearities of cavity field detuning and laser power. This suggests that using cavity field detuning and input laser power, we can create a programmable optical switch.

## 3.1 Effect of external mechanical driving fields on optical bistability

In the previous numerical simulations, we studied the behavior of bistable (bifurcation) curve of the steady-state photon number in the absence of external mechanical driving fields, i.e., $\varepsilon_1$ and $\varepsilon_2$. Now, we introduce external mechanical pumps to drive the mirrors $MR_1$ and $MR_2$ and check for the abrupt changes in the bistability of the system. So, we drive the moving mirrors $MR_s$ with a mechanical pump $\varepsilon_j = \epsilon_j e^{-i\phi_1}$ $(j = 1, 2)$ and explore its effect on the intracavity photon number $|c_s|^2$.

When a mechanical pump field is applied to MR$_1$ only, by keeping the amplitude $\epsilon_1$ constant and varying the phase angle $\phi_1$, it leads to phase-sensitive optical behaviors of the NEOMS. This behavior is shown in Fig. 6(a). When we increase the phase angle $\phi_1$ from $\pi/4$ to $\pi$, the steady-state intracavity photon intensity is amplified (see purple, purple dashed amd red dot-dashed curves). The width of bistable (bifurcation) curve also increases by increasing the phase angle $\phi_1$ of mechanical pump $\varepsilon_1$ on MR$_1$. An additional effect is also seen in the plot of bistability i.e., the curves overlap and the intersection point of the three different curves are same (see Fig. 6(a)). Moreover, the first upper stable branches follow the next upper stable path and vice versa. A similar transition occurs in the lower stable branches. Now, if we drive the second moving mirror MR$_2$ only, i.e., $\varepsilon_2 \neq 0$ and $\varepsilon_1 = 0$, keeping the amplitude $\epsilon_2$ constant and varying the phase angle $\phi_2$, the bistable behavior of the system confronts with some minor changes in the upper stable branch of the curve, as shown in Fig. 6(b). The first upper stable path follows the second upper stable path and vice versa, as we increase the phase angle $\phi_2$ from $\pi/4$ to $\pi$ (see blue solid, blue dashed and black long-dashed curves). In this case, the intracavity photon intensity is suppressed by increasing the phase angle $\phi_2$.

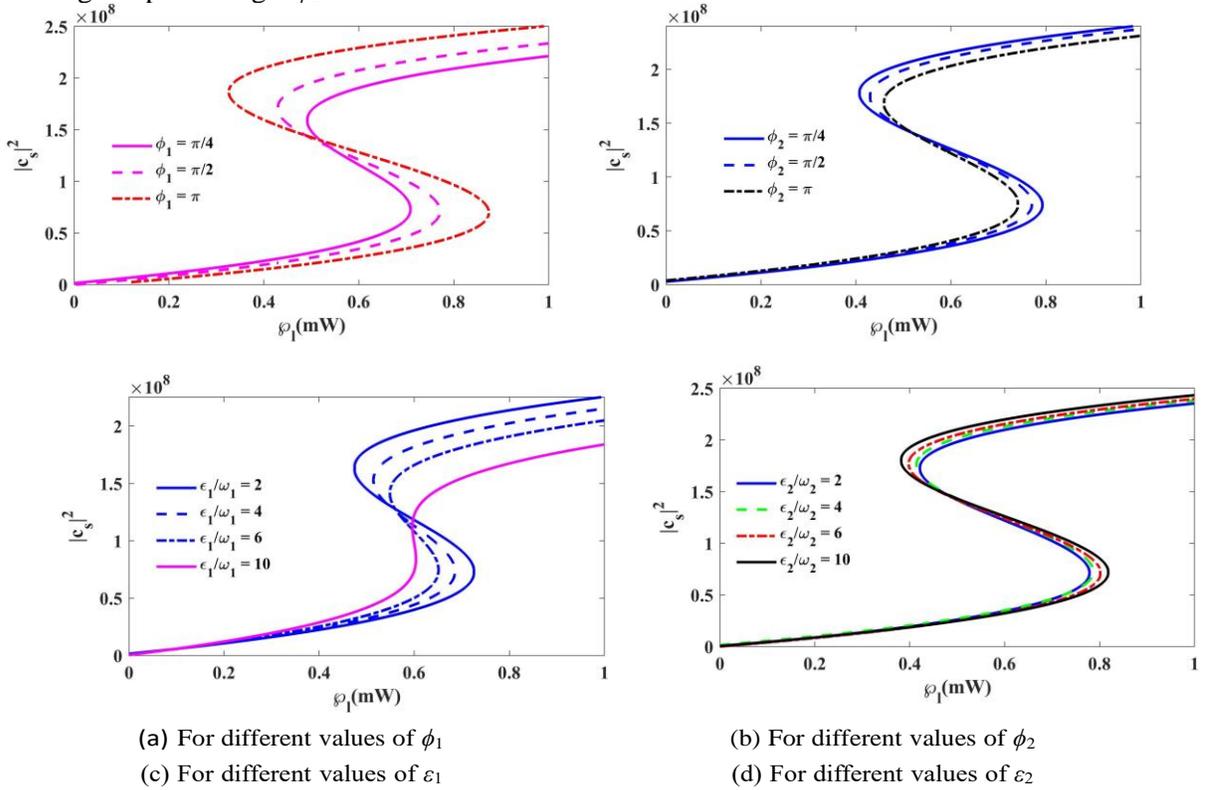

(a) For different values of $\phi_1$
(b) For different values of $\phi_2$
(c) For different values of $\varepsilon_1$
(d) For different values of $\varepsilon_2$

Figure 6: Plots of mean intra-cavity photon number $|c_s|^2$ versus the driving laser power $\wp_l$. Controlled bistable behavior for different values of amplitudes and phase angles of driving fields of MR$_1$ and MR$_2$. Other system parameters are same as used in Fig. (2) and Fig. (3).

Moreover, the same transition does not occur in the lower stable branches. The lower stable path is independent of the phase angle $\phi_2$. The main reason behind such a feature is described as: external modulating field on MR$_2$ only modifies the effective Coulomb coupling strength between the mirrors that has no any direct effect on the radiation pressure force. Similar behaviors of the bistable curve can be seen in Fig. 6(c) and Fig. 6(d) by selectively driving the moving mirrors MR$_1$ and MR$_2$, keeping the phase angle $\phi_1(\phi_2)$ constant and varying the amplitude $\epsilon_1(\epsilon_2)$ of MR$_s$.

## 4 Mirror Bi-Stability

In order to determine the bistable behavior of steady-state position $q_{1s}$ as a function of input laser power $\wp_l$, we use the relation $(b_{1s}^* + b_{1s}) = \sqrt{\frac{2m_1\omega_1}{\hbar}} q_{1s}$ and rewrite the Eq. (9) in terms of $q_{1s}$ as follows:

$$|\varepsilon_l + \varepsilon_p e^{-i\delta t}|^2 = |c_s|^2 \left[ \frac{\kappa^2}{4} + \left( \Delta_c - \sqrt{\frac{2m_1\omega_1}{\hbar}} G_0 q_{1s} \right)^2 \right] \quad (15)$$

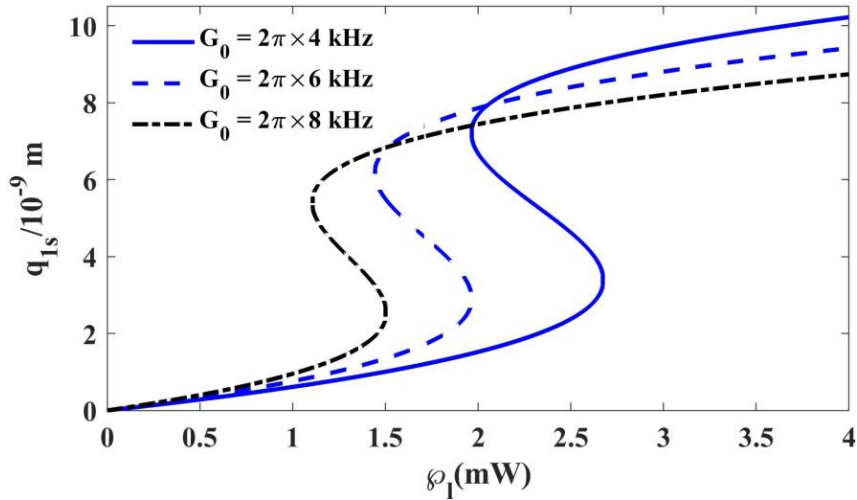

Figure 7: Plot of steady state displacement $q_{1s}$ versus the driving laser field power $\wp_l$ by varying the optomechanical coupling strength $G_0$. Other system parameters used are m$_1$ = m$_2$ = 145 ng, $\omega_1 = \omega_2 = 2\pi \times 947$ kHz, $\varepsilon_1 = \varepsilon_2 = 0, G_c = 0, \Delta_c = 3.6\kappa$ and $\kappa = 2\pi \times 215$ kHz.

The bistable behavior of steady-state position $q_{1s}$ as a function of input laser power $\wp_l$ is depicted in Fig. (7) for different values of optomechanical coupling strength $G_0$. In a weak coupling regime i.e., $G_0/2\pi = 4$ kHz (blue solid curve), the bistability in the system appears at a higher value of input laser power $\wp_l = 2.4$ mW. However, when the system is driven in a strong optomechanical coupling regime i.e., $G_0/2\pi = 6, 8$ kHz (see blue dashed and black dot-dashed curves), the bistability occurs at lower values of input laser power and the lower stable branch jumps to the upper stable path and continues to follow it, as the input laser power is increased. Moreover, the bistability curves overlap for higher values of laser power. This feature gives a control over the mechanical motion of micro-mirror MR$_1$ for different values of input laser power.

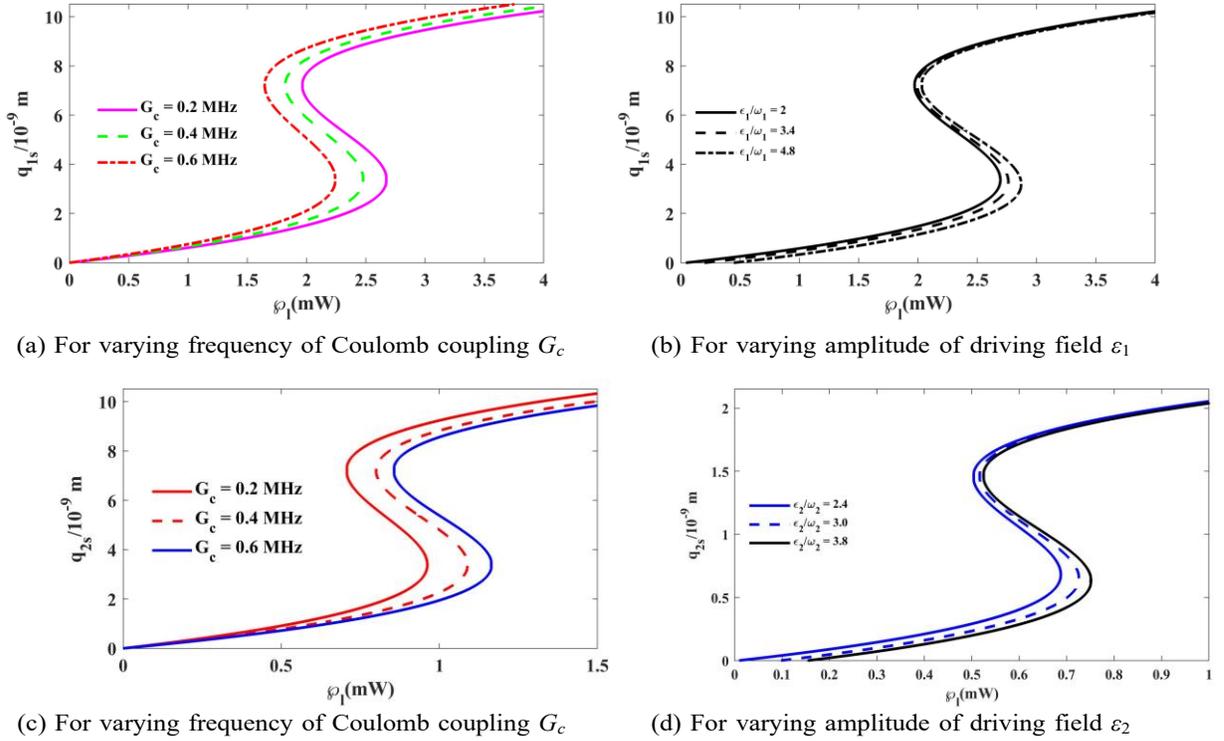

Figure 8: Plots of steady state displacement $q_{1s}$ versus the driving laser power $\wp_l$ by varying the Coulomb coupling strength $G_c$ and amplitude $\epsilon_1(\epsilon_2)$ of driving field $\varepsilon_1(\varepsilon_2)$ on MR$_1$(MR$_2$). Other system parameters are same as used in Fig. (7).

Now, we investigate the effect of Coulomb coupling strength $G_c$ and external mechanical pump field $\varepsilon_1$ on the steady-state position $q_{1s}$. For this reason, we plot the steady-state displacement $q_{1s}$ as a function of input laser power $\wp_l$ for several values of Coulomb coupling strength $G_c$ in Fig. 8(a). The steady-state solution shows a bistable (bifurcation) curve at a higher value of laser power for $G_c = 0.2$ MHz (see magenta solid curve). As we increase the bias gate V$_1$ on the mirror MR$_1$, the curve shifts to the left and the bistability occurs at lower values of laser power (see red dashed and red solid curves). Furthermore, no overlapping occurs and the first lower stable path follows the second stable branch and vice versa. Similar transitions can be seen in the upper stable branches. When an external modulating field $\varepsilon_1$ is applied on the mirror MR$_1$ only by keeping the phase angle $\phi_1 = 0$ and varying the amplitude $\epsilon_1/\omega_1 = 2, 3.4, 4.8$, the first lower stable path follows the second stable path and it continues to follow the third stable branch and so on (see black solid, dashed and dot-dashed curves in Fig. 8(b)). The upper stable branches do not exhibit any changes by varying the amplitude of external mechanical pump field on MR$_1$.

To study the effects of Coulomb force and mechanical pump on MR$_2$, we solved Eq. (7) for the steady-state displacement $q_{2s}$. The plot of steady-state solution for $q_{2s}$ shows a bistable behavior which is depicted in Fig. 8(c) and 8(d). The upper stable branch of the red solid curve ($G_c = 0.2$ MHz) jumps down to the upper stable path of red dashed curve ($G_c = 0.4$ MHz) and vice versa. In this case the bistability curve produces a reverse effect of Fig. 8(a). When an external mechanical driving field $\varepsilon_2$ is applied on MR$_2$ only, the system shows slight changes in the stable points of the bifurcation curve by increasing the amplitude of driving field $\epsilon_2/\omega_2 = 2.4, 3.0, 3.8$ and keeping the phase angle $\phi_2 = 0$ (see Fig. 8(d)). Furthermore, amplitude of steady-state displacement q$_{2s}$ decreases for the lower stable branch as we increase the amplitude of external modulating field $\varepsilon_2$ on MR$_2$. From the above discussions, it is cleared that a controllable bistable optical switching of steady-state photon intensity and mechanical displacements can be achieved by adjusting the coupling frequencies and selectively driving the moving mirrors MR$_1$ and MR$_2$.

# 5  Conclusion

We present a powerful scheme to experimentally realize optical switches, mainly based on the optical bistability of a nano-electro-opto-mechanical system (NEOMS) which is a compound system consisting of an optical resonator and two moving mirrors coupled by the electrostatic Coulomb force. We report optical and mirror bistability as a function of input laser power, coupling frequencies, and external mechanical pumping on the mirrors. We investigate the robustness of the system and provide dynamic control of the optical cavity field by mechanical movement of the mirror. The reverse is also true. We elaborate on the results by adjusting the amplitude and phase angle to selectively drive the mechanically moving mirrors. First, we examine the compound system constructed for optical bistability at steady-state photon intensities by adjusting the optomechanical, Coulomb coupling and cavity detuning frequencies. By adjusting the system to different optomechanical, Coulomb coupling and cavity detuning frequencies, we find that we can control the steady-state photon bistability by scanning the system from low or high value of input laser power. We report controlling the optical bistability of the system in the presence of two external mechanical driving fields on the mirrors. It can be seen that by selectively driving one of the mechanical pumps, the magnitude of the steady-state photon count can be increased and the optical bistability can be controlled by improving the mechanical motion of the mirror. We drive the mirrors by changing the amplitude and phase of the external mechanical driving fields, indicating that the bistability of static photon counts is phase sensitive. Moreover, we report the existence of mirror bistability as a function of optomechanical coupling and Coulomb coupling strengths. The stable branches at the top of the bistable curves overlap at different mirror-field coupling frequencies. In the presence of external driving fields on the moving mirrors, the mirror bistability indicates suppression and enhancement of the mirror's steady-state displacement. In the studied system, the coupling frequency and switchable mechanical driving fields allow detuning of the cavity field and threshold laser power that can be used to develop adjustable optical switches and all-optical transistors. This work can be extended to achieve multistable behaviour in both static photon intensity and mirror displacement.

# Acknowledgement

FS thanks UEC, Tokyo where a part of the work was performed. He also thanks JSPS and HEC (through NRPU grant#16427) for financial support.